\documentclass[preprint]{ptephy_v1}

\preprintnumber{XXXX-XXXX} 
\usepackage{hyperref}
\newcommand{\ttbar}{${t\bar{t}}~$}
\begin{document}

\title{B-Jet Tagging with Retentive Networks: A Novel Approach and Comparative Study}

\author[1]{Ayse Asu Guvenli}
\affil{Physics Department, Ozyegin University}

\author{Bora Isildak}
\affil{Physics Department, Yildiz Technical University}


\begin{abstract}

Identifying jets originating from bottom quarks is crucial for collider experiments aimed at discovering new physics. This paper presents a novel approach for b-jet tagging using Retentive Networks (RetNet), which leverages both low-level features of jet constituents and high-level jet features. The study utilizes a simulated \ttbar dataset provided by the CERN CMS Open Data Portal, focusing on only the semileptonic decays of \ttbar pairs produced from 13 TeV proton-proton collisions. 

We compare the performance of the proposed Retentive Network model against state-of-the-art models such as DeepJet and Particle Transformer, as well as a baseline Multi-Layer Perceptron (MLP) classifier. The Retentive Networks exhibit promising performance despite being trained on a relatively smaller dataset.
\end{abstract}

\subjectindex{Machine Learning, High Energy Physics}

\maketitle

\section{Introduction}

Jets, collimated sprays of particles, are among the most frequently observed phenomena in hadron collider experiments. Determining the parton origin of a jet is a fundamental and challenging task in this field. Many searches for new physics that aim to probe potential beyond the Standard Model phenomena require precise identification of the quark type that initiates jet formation.

B-tagging methods have evolved from simple cut-based models to machine learning-based approaches. Machine learning-based approaches for b-tagging can be broadly divided into two categories. The first category includes traditional machine learning techniques such as Naive Bayes classifiers\cite{CMS-PAS-BTV-09-001, collaboration_2013}, perceptrons \cite{CMS-PAS-BTV-15-001, 2016_atlas}, and decision trees \cite{CMS-PAS-BTV-15-001}. The second category comprises methods utilizing deep neural networks (DNNs)\cite{Sirunyan_2018}, such as Convolutional Neural Networks (CNNs), Recurrent Neural Networks, and Long Short-Term Memory (LSTM) networks {LSTM}.

In recent years, the ATLAS b-tagging algorithm, based on the concept of Deep Sets \cite{ATL-PHYS-PUB-2020-014}, has mapped input trace features to a latent space and then combined them with a simple per-feature aggregation. While this approach shows comparable performance to the RNN-based method \cite{ATL-PHYS-PUB-2017-003}, the training and inference times are lower for the same inputs.

For over five years, the CMS Collaboration has employed a fully connected neural network called DeepCSV for b-tagging \cite{DeepCSV}, which utilizes both the track and secondary vertex (SV) information.

A CNN-based approach called DeepJet\cite{DeepJet}, which uses around 650 input features, has also been introduced by the CMS Collaboration, outperforming existing b-jet taggers. Later, ParticleNet \cite{Qu_2020}, a graph neural network (GNN) based approach using particle clouds to represent jets, replaced DeepJet. In recent years, with the rise of the use of attention mechanisms, the particle transformer (ParT)\cite{qu2024particletransformerjettagging} has replaced state-of-the-art models. As the name suggests, ParT is a transformer-based architecture that takes the kinematics of particles, particle identification (PID) information, and trajectory displacements and then feeds them through an attention mechanism. These state-of-the-art models achieve superior performance on this task but require extensive datasets and high computational power for the training.

This paper proposes an alternative approach for b-jet tagging that leverages the retention mechanism \cite{retentive}. Which treats particle track, SV, and global jet features separately.
\newpage
\section{Data Simulation and Input Data Pre-processing}
\label{sec: Data}

\subsection{Data Source and Simulation}
The model is trained on approximately four million jets obtained from a simulated semileptonic \ttbar dataset that is available on the CERN CMS Open Data portal \cite{CERN_OpenData_2012}. It corresponds to proton-proton collisions at a center-of-mass energy of 13 TeV, with 2016 CMS data-taking conditions. The dataset was produced through a multi-step simulation process using the \textsc{Powheg} generator for next-to-leading-order (NLO) hard scattering, followed by parton showering and hadronization with \textsc{Pythia8}. 

The simulation workflow includes particle propagation through the CMS detector simulation afterward until the final high-level reconstructed objects are obtained. The jets are clustered using the anti$-k_t$ algorithm \cite{Cacciari_2008} with cone size parameter $\Delta R=0.4$. $p_T > 30$ GeV and $|\eta| < 2.5$ cuts are applied to each jet. 

The hadron flavor information available in the simulated data is used as true jet flavor labels, which is the jet's ghost clustered flavor \cite{Sirunyan_2018}. Jets are separated into three categories: b-jets (from b-quarks), c-jets (from c-quarks), and light jets (from gluons or u, d, s quarks).

\subsection{Jet Feature Extraction}
Three different levels of information are used for a given jet: global jet features, track features, and SV features. Recent studies suggest that low-level information could improve the performance of jet classification problems \cite{paganini2017}; hence, only low-level inputs were used. 

\begin{figure}[h!]
    \centering
    \includegraphics[width=1\linewidth]{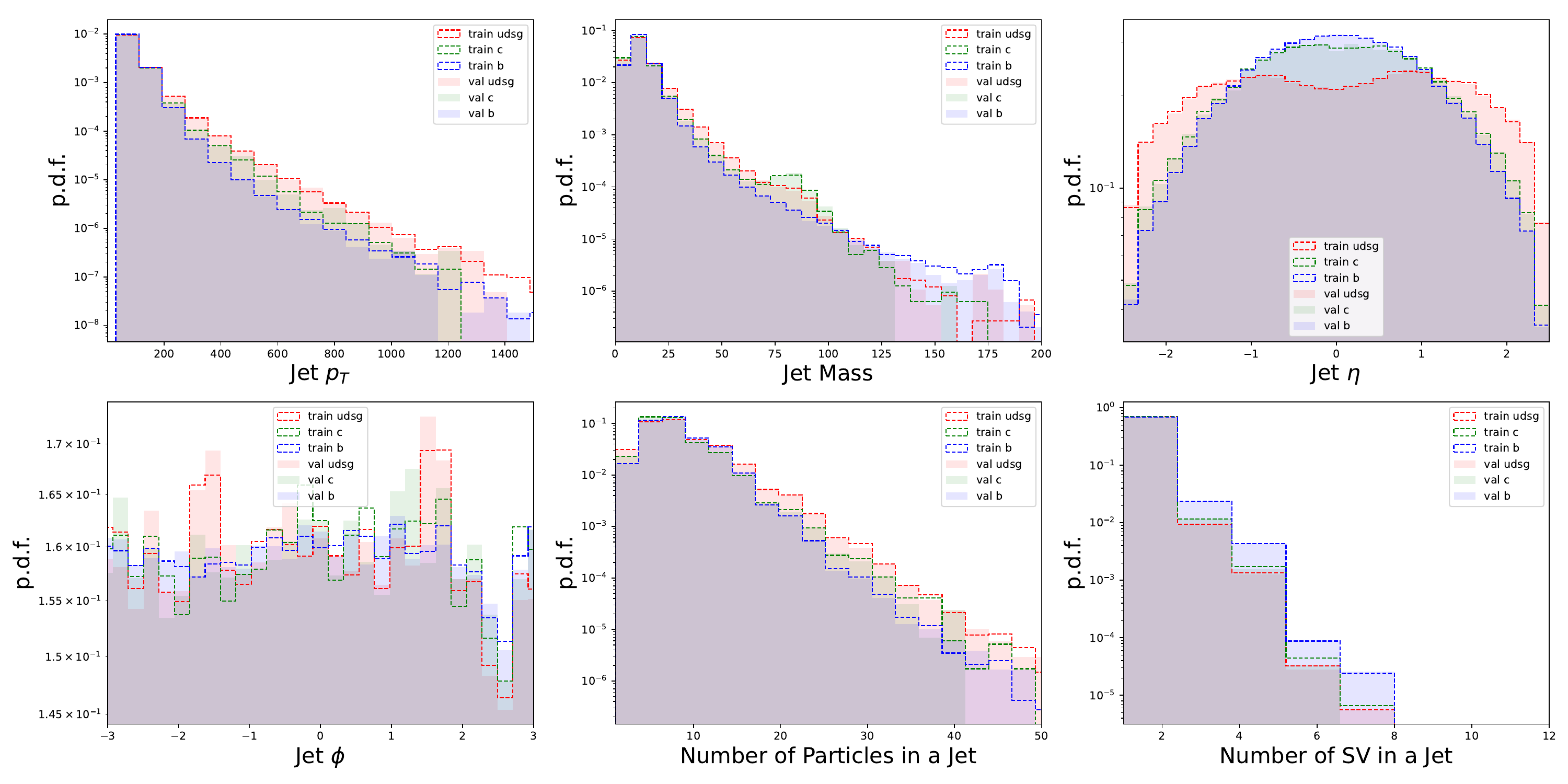}
    \caption{Global jet feature histograms for b, c, and light jets.}
    \label{fig:jet_histo}
\end{figure}

For the global jet features, jet $p_T$, jet mass, jet $\eta$, jet $\phi$, the number of tracks, and the number of SVs in the jet were selected. The probability density functions for each jet feature are shown in Fig.\ref{fig:jet_histo}. Here, to understand the discriminative power of each feature, the aforementioned sub-classes are shown separately. Training and validation sets are shown separately as well in order to identify any potential biases.

\begin{figure}[h!]
    \centering
    \includegraphics[width=1\linewidth]{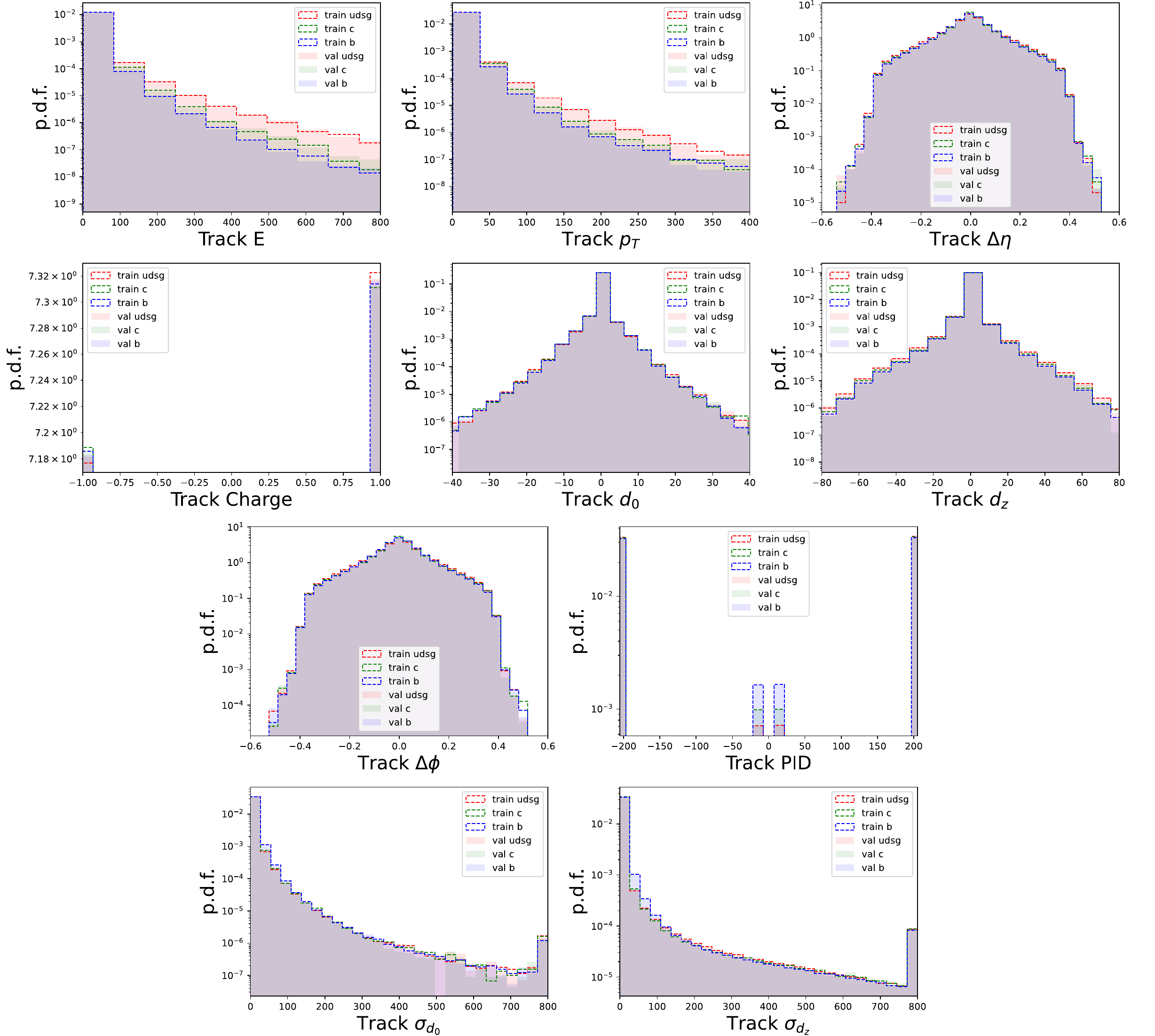}
    \caption{Track feature histograms for b, c, and light jets.}
    \label{fig:track_histo}
\end{figure}

Another cut for the track features was applied by selecting tracks in jets, with $p_T$ greater than $1.0$ GeV and a normalized $\chi^2$ less than $5.0$. The selected track features are: track $p_T$, track energy, track charge, track PID, transverse impact parameter ($d_0$), longitudinal impact parameter $d_z$, track $d_0$ significance ($\sigma_{d_0}$), track $d_z$ significance ($\sigma_{d_z}$), $\Delta\eta$ and $\Delta\phi$ with respect to jet $\eta$, jet $\phi$. The probability density functions of track features are given in Fig.\ref{fig:track_histo}.

\begin{figure}[h!]
    \centering
    \includegraphics[width=1\linewidth]{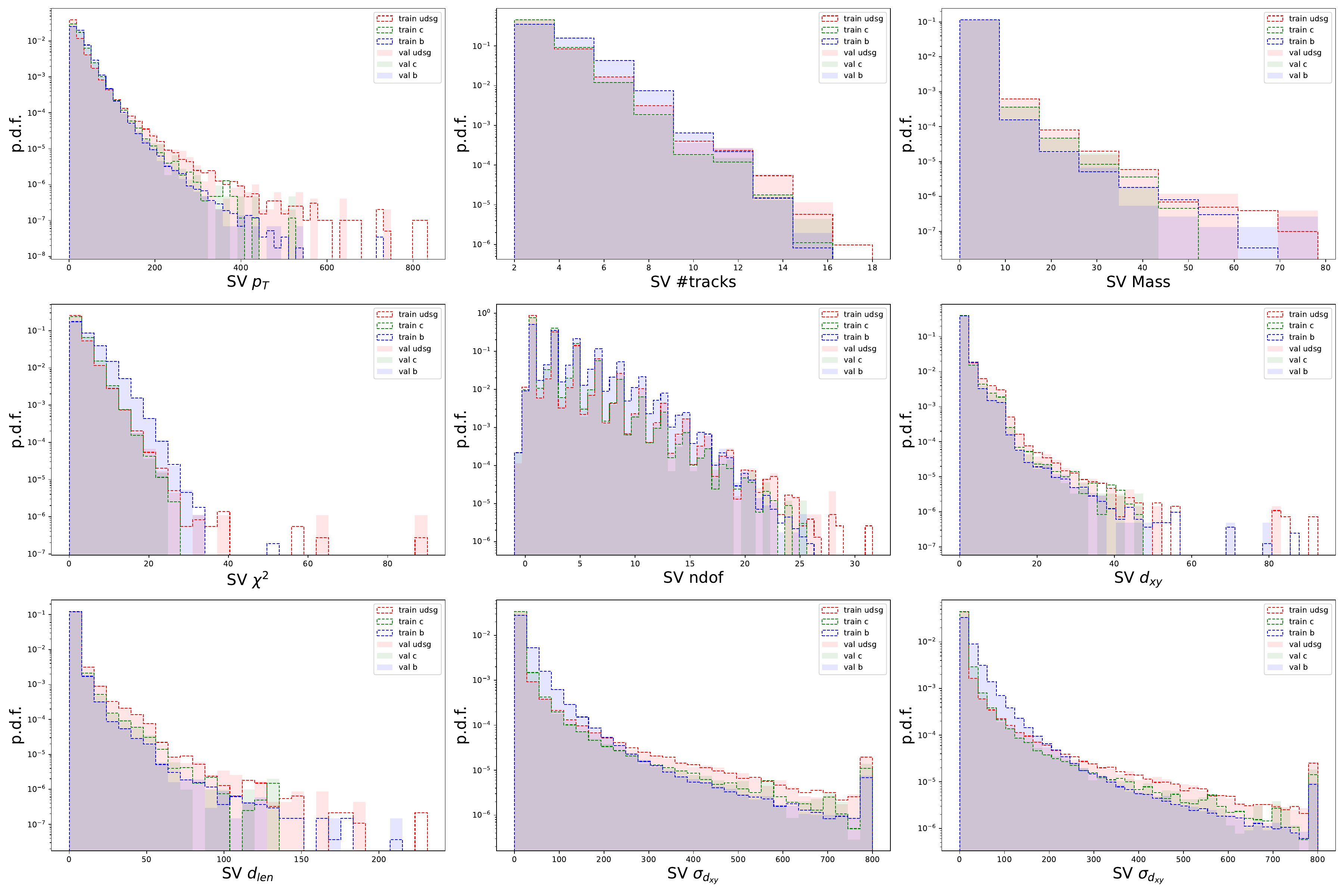}
    \caption{Secondary Vertex feature histograms for b, c, and light jets.}
    \label{fig:sv_histo}
\end{figure}

\begin{figure}[h!]
    \centering
    \includegraphics[width=0.5\linewidth]{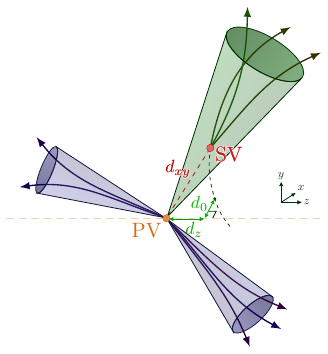}
    \caption{Schematic showing track and vertex impact parameters. The image was generated by using the script provided by \cite{jet_btag_tikz}.}
    \label{fig:impact}
\end{figure}

Each SV is matched to the jet by checking if it is within $\Delta R<0.4$ of the jet, then the following information is obtained: SV $p_T$, SV mass, SV $\chi^2$, number of tracks associated to the SV, SV transverse impact parameter ($d_{xy}$), SV 3-D impact parameter ($d_{len}$), SV $d_{xy}$ significance ($\sigma_{d_{xy}}$), and SV $d_{len}$ significance ($\sigma_{d_{len}}$) as shown in Fig.\ref{fig:sv_histo}. An illustrative schematic of track and vertex impact parameters are given in Figure \ref{fig:impact}.

\subsection{Data Preprocessing}
After deriving these features, each jet's track and SV features are sorted in decreasing $p_T$ order. This approach serves as a proxy for the time order of jet formation, as the true time-order information is experimentally inaccessible. 

Each jet uses a fixed number of 16 tracks and 5 SVs. This is done by clipping or padding each feature with zero values. This step is necessary, as most deep-learning models are not structured to work with variable-size inputs.

\begin{figure}[h!]
    \centering
    \includegraphics[width=1\linewidth]{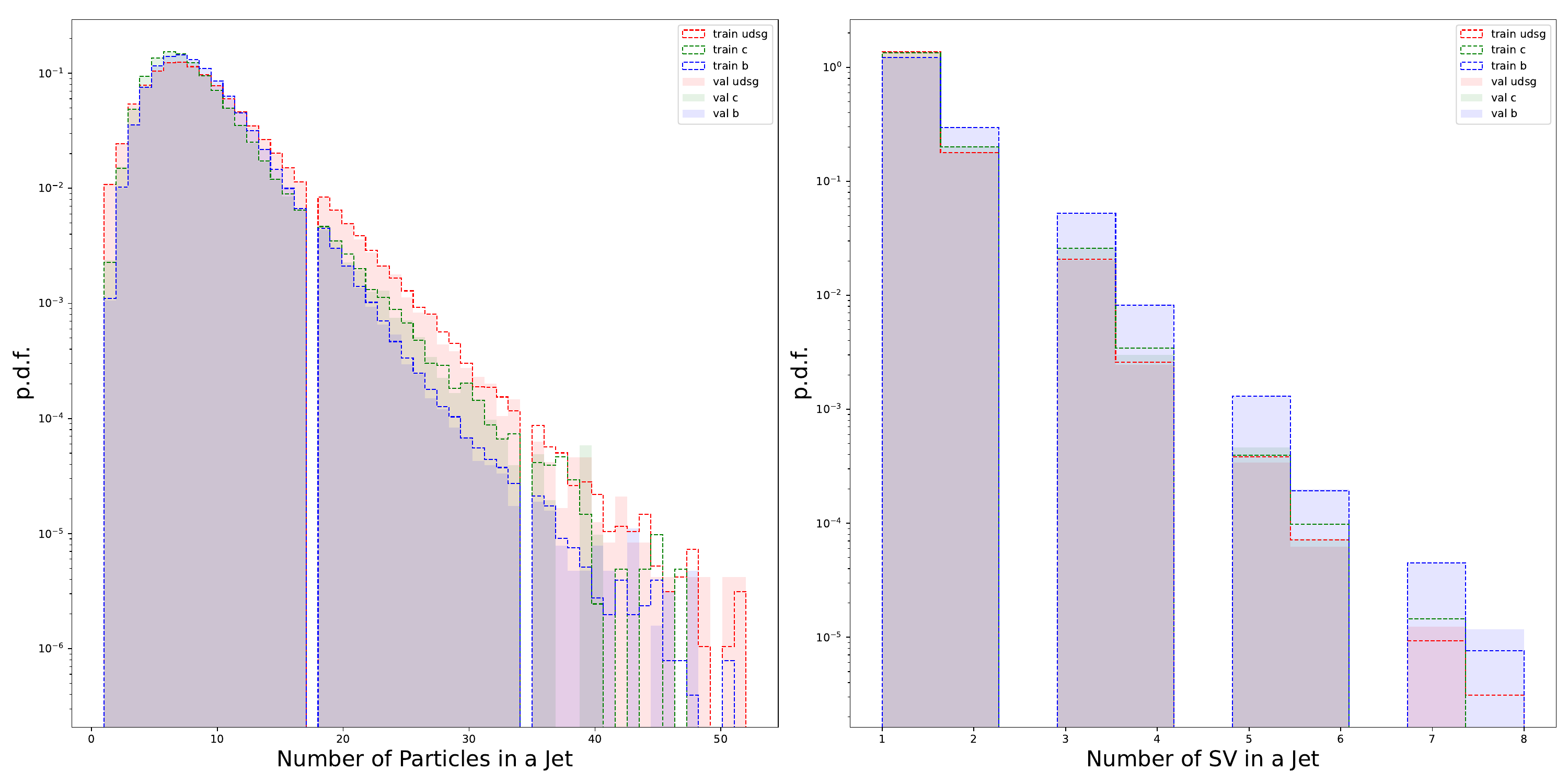}
    \caption{Number tracks (right) and the number of SV's (left) in a jet.}
    \label{fig:cut_len}
\end{figure}

The percentage of jets in our dataset with less than 17 tracks is $\% 2.63$, and the percentage of jets in our dataset with less than 6 SV is $\% 0.012 $. Also, the percentage of b-jets with less than 17 tracks is $\% 1.99$, and the percentage of b-jets in our dataset with less than 6 SV is $\% 0.016$. We still keep these jets in our dataset but set a limit to the maximum number of tracks and SVs. Probability density functions for the number of tracks and SVs in each jet are shown in Fig.\ref{fig:cut_len}.

After the cuts, we were left with a sample of 5 million jets: 4 million were used for training, and 1 million were used for validation. The hadron flavor label is converted to binary labels by assigning a value of one to b quark jets and a value of zero to u, d, s, c quark jets, and gluon jets.

\newpage
\section{Methods}
\label{sec: Methods}

\subsection{Retention Mechanism}
\label{sec: Retentive Networks}

Retentive Networks (RetNet) are designed to handle sequential data by selectively retaining important information from previous inputs\cite{retentive}. Unlike recurrent networks, which store information in hidden states, RetNet employs a retention mechanism that enables efficient processing of sequential inputs without recurrence. This makes it particularly well-suited for jet data, where the structure of jet constituents (tracks, SVs) reflects complex sequential patterns.

The RetNet architecture incorporates a mechanism conceptually similar to attention, which is called the retention mechanism and can be summarized in Equations \ref{eq:retention} and \ref{eq:retention2}. Here, as in the attention mechanism \cite{vaswani2023attentionneed}, we have Key (K), Querry (Q), and Value (V) matrices with the learned parameters $W_Q, W_K, W_V$, and input $X$. The significant difference arises from $\Theta$ and $D$ matrices, whose role is to introduce exponential decay and causal masking.

\begin{equation}
    \begin{aligned}
    Q &= (XW_Q) \odot \Theta, \quad K = (XW_K) \odot \overline{\Theta}, \quad V = XW_V, \\
    \Theta_n &= e^{i n \theta}, \quad
    D_{nm} = 
        \begin{cases} 
        \gamma^{n-m}, & n \geq m, \\
        0, & n < m,
        \end{cases}
    \end{aligned}
    \label{eq:retention}
\end{equation}

\begin{equation}
    \text{Retention}(X) = (QK^\top \odot D)V
    \label{eq:retention2}
\end{equation}

As attention is mainly used in multi-head attention modules, the retention mechanism is also embodied in a multi-scale retention (MSR) module. This module also has multiple heads, but each head's scale ($\gamma$) also changes, hence the name. Apart from the retention, MSR also includes the group-norm layer and a final swish gate to increase non-linearity. The principle of MSR is given in Equation \ref{eq:msr}.

\begin{equation}
    \begin{aligned}
    \gamma &= 1 - 2^{-5 - \text{arange}(0, h)} \in \mathbb{R}^h, \\
    \text{head}_i &= \text{Retention}(X, \gamma_i), \\
    Y &= \text{GroupNorm}_h\big(\text{Concat}(\text{head}_1, \dots, \text{head}_h)\big), \\
    \text{MSR}(X) &= \big(\text{swish}(XW_G) \odot Y\big)W_O
    \end{aligned}   
    \label{eq:msr}
\end{equation}

Where $W_G, W_O \in \mathbb{R}^{d_{\text{model}} \times d_{\text{model}}}$ are learnable parameters and $h$ is the number of heads. Each RetNet block has an MSR module and a Feed-forward Network (FFN) layer.

\subsection{JetRetNet}

\begin{figure}[h!]
    \centering
    \includegraphics[width=1\textwidth]{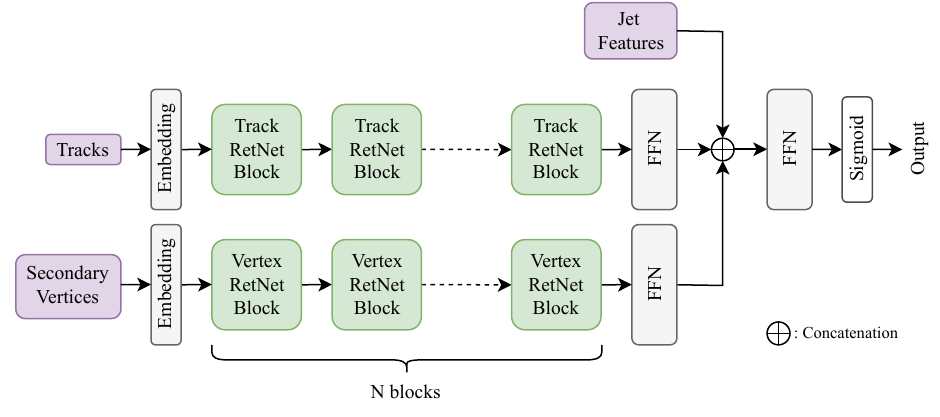}
    \caption{Architecture of the JetRetNet model.}
    \label{fig:JetRetNet}
\end{figure}

Our model, JetRetNet, given in Figure\ref{fig:JetRetNet}, is designed to process three key types of input features: Track, SV, and Global jet features. Track and SV features are first passed through an embedding layer. Then, they are fed into a separate series of RetNet blocks designed to capture sequential dependencies between the tracks and the SVs in parallel.

The outputs from both RetNet modules are given to FFNs and then concatenated along with the global jet features. The concatenated features are passed through an FFN layer, serving as the final b-jet tagging classifier. The FFN consists of fully connected layers with non-linear activation functions, followed by a sigmoid output layer.

\subsection{Model Implementation}
\label{sec:Model Implementation}

The Retentive Network (RetNet) model was implemented in PyTorch \cite{paszke2019pytorchimperativestylehighperformance} using two independent RetNet modules \cite{retentive}.

A batch size of 512 was employed during training, along with the Adam optimizer initialized with a learning rate of $0.0005$. The learning rate was reduced by $10\%$ at each epoch. The model that achieved the best performance included an embedding layer with 128 neurons and a dropout layer with a rate of 0.1. The architecture incorporated four layers of RetNet blocks with a hidden dimension of 32 and 8 heads, two feedforward networks (FFNs) with two layers of 128 neurons to process outputs from the retention layers for tracks and secondary vertices (SVs), and a final FFN with three layers of 128 neurons. All FFN layers utilized dropout rates of 0.1 to mitigate overfitting.

To compare performance, the RetNet-based model was benchmarked against an MLP model configured with the same hyperparameters, including four layers and a hidden dimension of 128.

Both models were trained using an early-stopping algorithm that monitored improvements in loss. The training was terminated if the loss did not decrease by more than 0.003 within five epochs.

All training was conducted on two parallel NVIDIA Quadro RTX 5000 GPUs. Further fine-tuning of the models remains possible.

\newpage
\section{Results and Conclusion}
\label{sec:Results}

\begin{figure}[h!]
    \centering
    \includegraphics[width=1.0\linewidth]{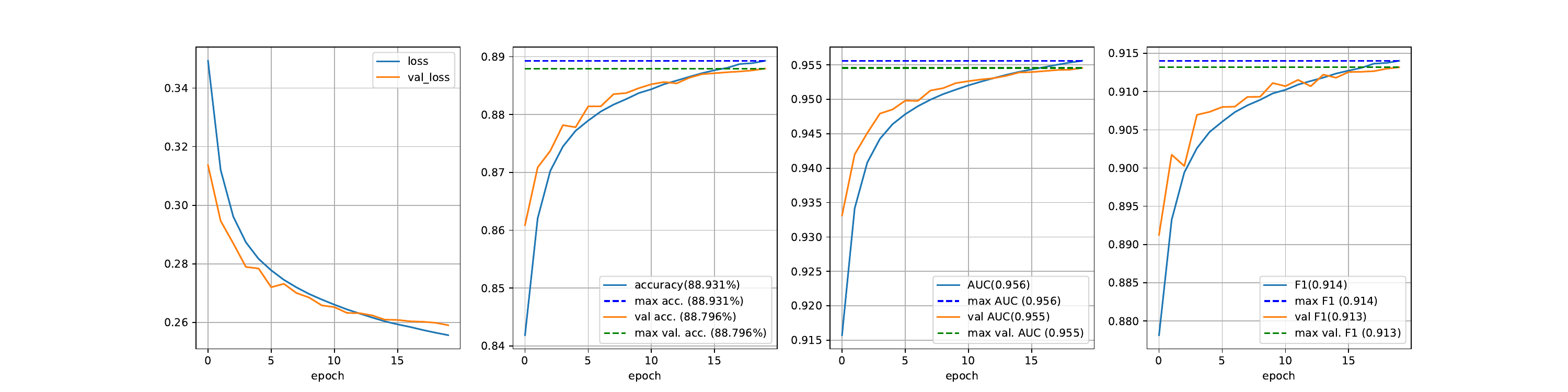}
    \caption{Evolution of loss, accuracy, AUC, and F1-score of JetRetNet during training. }
    \label{fig:train}
\end{figure}

The JetRetNet model is trained for 20 epochs using an early stopping condition mentioned in the previous section. Figure \ref{fig:train} shows the loss, accuracy, Area Under the ROC Curve (AUC), and F1-score evolutions through epochs. Among those, accuracy cannot be considered a decisive metric since it highly depends on the class imbalance in the dataset. Yet, the AUC and F1-score objectively show that the model learned to discriminate b-jets from light parton-originated jets (u, d, s, c, g). We also followed the same procedure for training an MLP for performance comparison.

\begin{figure}[h!]
    \centering
    \includegraphics[width=1\linewidth]{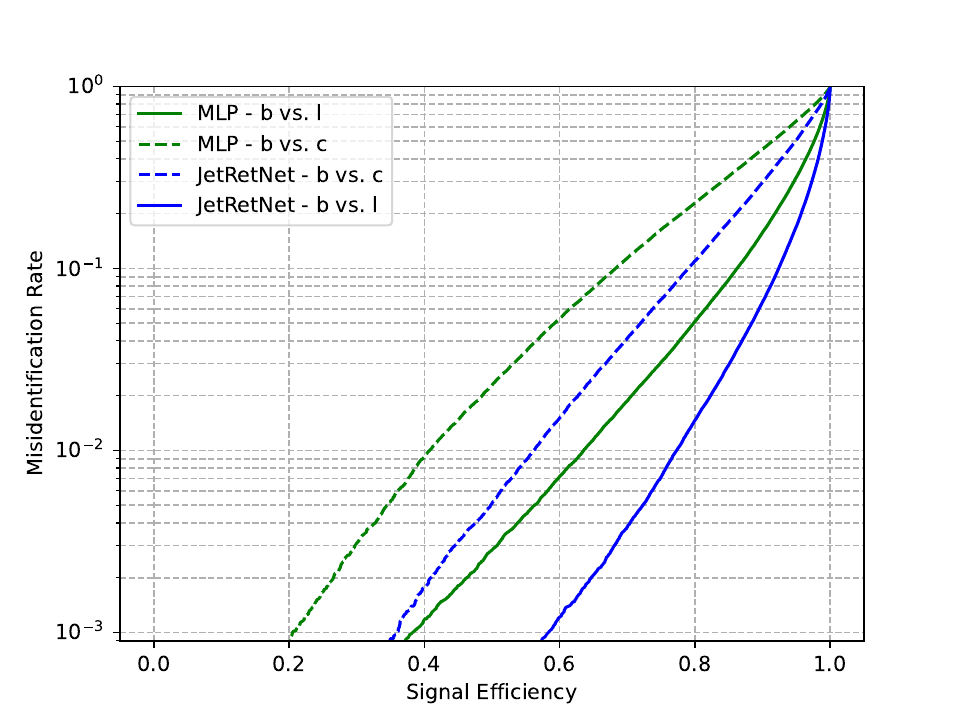}
    \caption{Mis-identification rate vs. signal efficiency, a fair comparison of MLP and JetRetNet }
    \label{fig:ROC_fair}
\end{figure}

Looking at the ROC plot provides a better understanding of the model's performance. The misidentification rate in terms of b-jet tagging efficiency for light jets and c-jets can be found in Figure \ref{fig:ROC_fair} separately. In the figure, the more the curve is right-skewed, the better the model's performance is. As can be seen in the figure, JetRetNet outperforms the MLP for all misidentification rates.

It's common practice to compare model performances by looking at the tagging efficiency for several fixed misidentification rates. In this study, we have considered three different misidentification rates, 0.1, 0.01, and 0.001, called light, medium, and tight, respectively, as working points.

\begin{figure}[h!]
    \centering
    \includegraphics[width=0.8\linewidth]{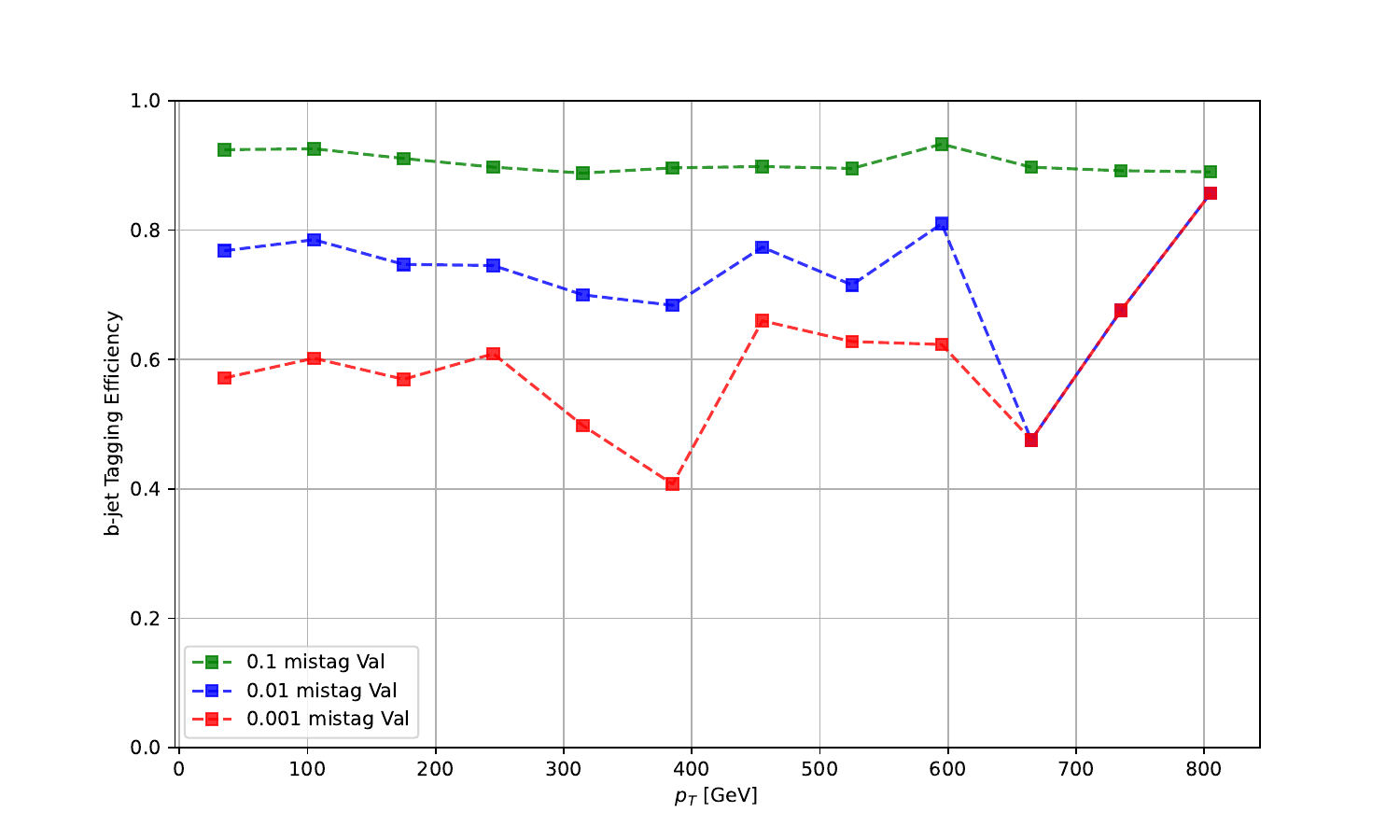}
    \caption{b-jet vs. light-jet classification efficiency of JetRetNet for $p_T$ ranges.}
    \label{fig:pt_l}
\end{figure}

\begin{figure}[h!]
    \centering
    \includegraphics[width=0.8\linewidth]{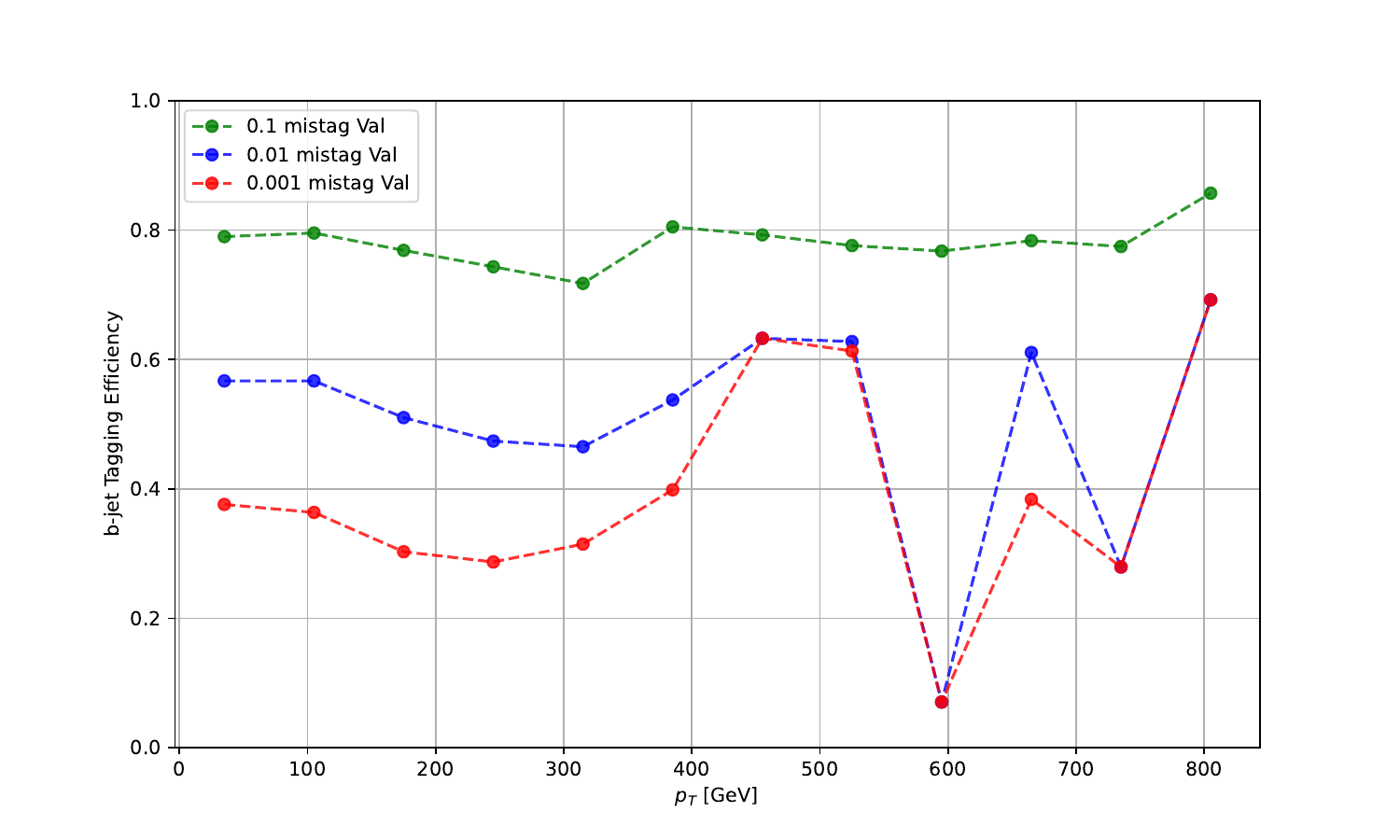}
    \caption{b-jet vs. c-jet classification efficiency of JetRetNet for $p_T$ ranges.}
    \label{fig:pt_c}
\end{figure}

A more detailed performance analysis can be done by inspecting the tagging efficiency for various $p_T$ ranges. In Figures \ref{fig:pt_l} and \ref{fig:pt_c}, b-jet tagging efficiency w.r.t. $p_T$, is shown for b-jets vs. light jets and b-jets vs. c-jets, respectively for the 3 working points. Also, both performances are given to make sure there are no discrepancies between the training and the validation datasets.

For b vs. c jet classification, we observe a distinct pattern in tagging efficiency across different $p_T$ intervals (Fig. \ref{fig:pt_c}). The JetRetNet model consistently outperforms the MLP.

\begin{figure}[h!]
    \centering
    \includegraphics[width=1\linewidth]{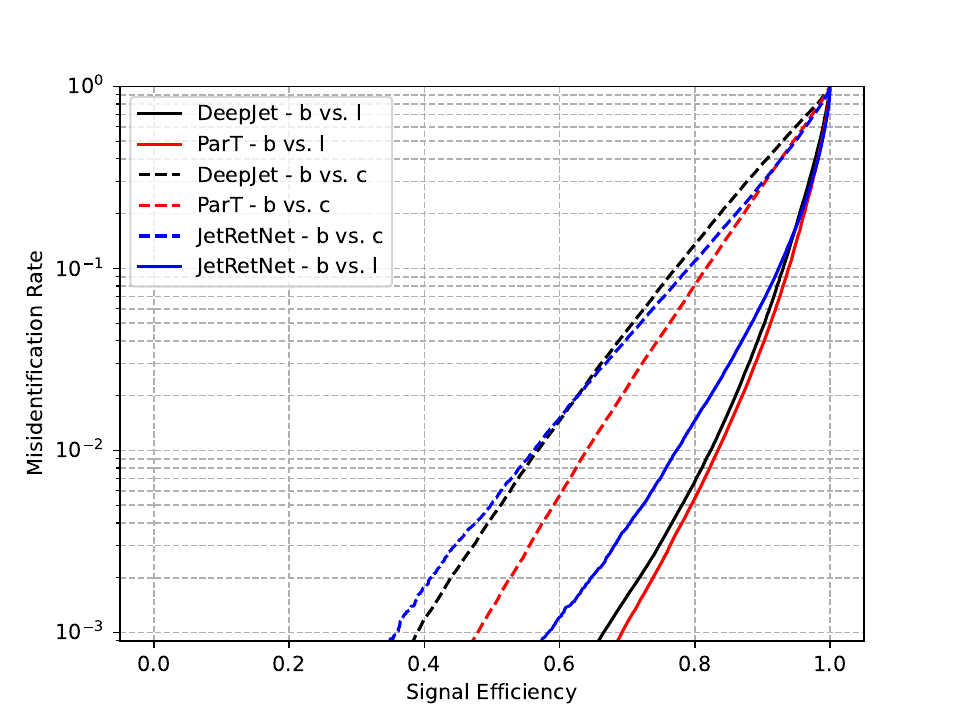}
    \caption{Mis-identification rate vs. signal efficiency, an unfair comparison of DeepJet, ParT, and JetRetNet.}
    \label{fig:ROC_unfair}
\end{figure}

Although it is not fair to compare JetRetNet's performance with state-of-the-art models like DeepJet and ParT, it is essential to understand its relative performance. In Figure \ref{fig:ROC_unfair} we give the performance comparison of JetRetNet with DeepJet and ParT. This comparison is unfair as we used a training dataset size two orders smaller than the dataset used in training DeepJet and ParT. JetRetNet cannot outperform these models in any of the working points but gives promising results, especially at tight WP. Hinting at its potential for future use.

In this paper, we explored the use of Retentive Networks for b-jet tagging and compared their performance to two state-of-the-art models, DeepJet and Particle Transformer. While our model did not outperform these larger models, we demonstrated that Retentive Networks could achieve competitive results, even with two orders of magnitude smaller datasets. This highlights the potential of Retentive Networks for experiments with limited computational resources or smaller datasets.

Future work will focus on scaling the dataset size, experimenting with more complex architectures, and exploring other applications of Retentive Networks in high-energy physics.

\subsection*{Acknowledgement}
The authors acknowledge the CERN Open Data portal and the CMS Collaboration for providing the publicly available datasets used in this study. 

\newpage
\vspace{0.2cm}
\noindent
\let\doi\relax

\bibliographystyle{unsrt}
\bibliography{Manuscript}

\end{document}